\documentclass{segabs}

\usepackage{graphicx}
\usepackage{amsfonts}
\usepackage{amsmath}
\usepackage{epstopdf}
\usepackage{bm}
\usepackage{caption}
\usepackage{subfig}
\usepackage{alphalph}
\usepackage{xcolor}
\usepackage{tikz}
\usepackage{pgfplots}
\usepackage{subfigure}
\usepackage{sansmath}

\usepackage{enumitem}

\begin{document}

\onecolumn 

\begin{description}[labelindent=-1.5cm,leftmargin=1cm,style=multiline]

\item[\textbf{Citation}]{Yazeed Alaudah, Shan Gao, and Ghassan AlRegib (2018) Learning to label seismic structures with deconvolution networks and weak labels. SEG Technical Program Expanded Abstracts 2018: pp. 2121-2125.}

\item[\textbf{DOI}]{https://doi.org/10.1190/segam2018-2997865.1}

\item[\textbf{Review}]{Date of publication: August 2018}

\item[\textbf{Data and Codes}]{https://ghassanalregib.com/publications/}

\item[\textbf{BibTex}] {
\begin{verbatim}
@inbook{doi:10.1190/segam2018-2997865.1,
author = {Yazeed Alaudah and Shan Gao and Ghassan AlRegib},
title = {Learning to label seismic structures with deconvolution networks and weak labels},
booktitle = {SEG Technical Program Expanded Abstracts 2018},
chapter = {},
pages = {2121-2125},
year = {2018},
doi = {10.1190/segam2018-2997865.1},
URL = {https://library.seg.org/doi/abs/10.1190/segam2018-2997865.1},
eprint = {https://library.seg.org/doi/pdf/10.1190/segam2018-2997865.1}
}
\end{verbatim}
} 

\item[\textbf{Contact}]{alaudah@gatech.edu  OR alregib@gatech.edu\\ http://ghassanalregib.com/ }
\end{description}

\thispagestyle{empty}
\newpage
\clearpage
\setcounter{page}{1}

\title{Learning to Label Seismic Structures with Deconvolution Networks and Weak Labels}

\author{Yazeed Alaudah, Shan Gao, and Ghassan AlRegib\\
\{alaudah,gaoshan427,alregib\}@gatech.edu\\ Center for Energy and Geo Processing (CeGP) at Georgia Tech and KFUPM}

\vspace{-5pt}

\maketitle

\begin{abstract}
Recently, there has been increasing interest in using deep learning techniques for various seismic interpretation tasks. However, unlike shallow machine learning models, deep learning models are often far more complex and can have hundreds of millions of free parameters. This not only means that large amounts of computational resources are needed to train these models, but more critically, they require vast amounts of labeled training data as well. 

In this work, we show how automatically-generated weak labels can be effectively used to overcome this problem and train powerful deep learning models for labeling seismic structures in large seismic volumes.  

To achieve this, we automatically generate thousands of weak labels and use them to train a deconvolutional network for labeling fault, salt dome, and chaotic regions within the Netherlands F3 block. Furthermore, we show how modifying the loss function to take into account the weak training labels helps reduce false positives in the labeling results. 

The benefit of this work is that it enables the effective training and deployment of deep learning models to various seismic interpretation tasks without requiring any manual labeling effort. 

We show excellent results on the Netherlands F3 block, and show how our model outperforms other baseline models.

\end{abstract}

\vspace{-15pt}
\section{Introduction}
\vspace{-5pt}

Many machine learning based methods have been proposed for various seismic interpretation tasks. Many of these techniques are used to extract or interpret localized seismic structures such as salt domes, faults, horizons, and so on. \cite{spm} provides a good overview of such methods\textemdash and new emerging techniques\textemdash from a signal processing perspective. Given all this interest in automating various seismic interpretation tasks, not much work has been done on labeling \textit{entire} seismic volumes based on their dominant structures. \cite{icip2016} presented an early attempt at this using similarity-based image retrieval and a support vector machine (SVM) classifier. Additionally, \cite{Charles2017MalenoV} recently published code for facies classification using a basic 5-layer convolutional neural network (CNN). 

In recent years, deep learning has witnessed great success in wide-ranging applications and has revolutionized the fields of machine learning and computer vision. This success was not only due to the growing use of powerful GPUs, or the advent of deep learning models that can learn their own hierarchical data representations; but also, and more importantly, the arrival of very large labeled datasets, such as  ImageNet \cite[]{russakovsky2015imagenet}. Deep learning models are often far more complex than traditional machine learning models and can have hundreds of millions of free parameters. This not only means that they need large amounts of computational resources to train these models, but more critically, they require vast amounts of labeled training data. Labeled data can be extremely costly and time-consuming to obtain. In practice, the high cost of acquiring labeled data is a critical bottleneck to the successful application of deep learning to many application domains. This bottleneck is especially true for the field of seismic interpretation, where very few \textit{labeled} datasets are freely available.

There have been quite a few techniques published recently that apply deep learning algorithms, such as CNNs, to seismic interpretation problems (e.g., \cite{waldeland2017salt,araya2017automated,huang2017scalable, di2018seismic}. However, all \emph{these methods require ``strong" labels\footnote{Here, ``strong" labels mean high-quality labels generated by a domain expert. This is opposed to automatically-obtained ``weak" labels that convey far less information than strong ones, and are usually much noisier and less accurate, but are much easier to obtain. Please see \cite[]{alaudah2017weakly} for a more detailed explanation.}} that are obtained by manual labeling from an interpreter. Manually-labeling data for training deep learning models can be as laborious and time-consuming as manual interpretation workflows. Furthermore, over-training a network on a relatively small amount of manually labeled data can easily lead to overfitting, and therefore poor generalization performance.

\begin{figure}[ht!]

\begin{center}
\begin{tabular}{c|c|c}
\texttt{chaotic} & \texttt{faults} & \texttt{salt dome} 
\\
\includegraphics[width=1.8cm]{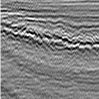} 
&
\includegraphics[width=1.8cm]{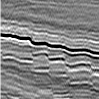} 
&
\includegraphics[width=1.8cm]{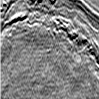} 
\\
\includegraphics[width=1.8cm]{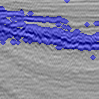}
&
\includegraphics[width=1.8cm]{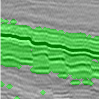}
&
\includegraphics[width=1.8cm]{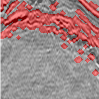}
\end{tabular}
\end{center}


  \caption{Sample results from our weakly-supervised label mapping technique in \cite[]{alaudah2017weakly}. The first row shows the original seismic images, with their image-level labels above. The second row shows the resulting pixel-level labels for these images. 
  }

\label{fig:img_labels}
\end{figure}

One solution is to use weakly-supervised methods that do not require manual labeling efforts and can produce orders of magnitude more labeled data than manual labeling by domain experts. However, these labels are usually of lower quality than manually obtained ones. In this paper, we build on our previous work on generating weak labels for seismic interpretation \cite[]{alfarraj_MMSP,alaudah2017weakly,alaudah2018geophysics}. Here, we show that despite their lower quality, weak labels are extremely useful in enabling the training of complex deep learning models\textemdash such as deconvolution networks\textemdash with no manual labeling effort required. The only requirement is for the interpreter to ``define" the various structures of interest to her by selecting at least one image for every seismic structure (as is shown in the \textit{first row} of Figure \ref{fig:img_labels}).

To summarize, our main contributions in this paper are: 
\vspace{-5pt}
\begin{itemize}
    \item We use automatically-generated weak labels to train deep learning models with no manual labeling effort required.
    \item We propose using a deconvolution network, trained on weak labels, to accurately localize and classify seismic structures.
    \item We modify the network's loss function to avoid putting too much trust in our weak labels, and therefore, reduce false positives.
\end{itemize}
\vspace{-5pt}
In the next section, we will explain these contributions in detail; and in the results section, we will show how these contributions help achieve excellent results on the Netherlands North Sea F3 block.

\section{Method}

\subsection{Weak labels}

In our previous work \cite[]{alfarraj_MMSP} we showed how seismic image similarity measures could be used to retrieve thousands of images based on their visual similarity. Using this technique, we can obtain thousands of image-level labeled seismic images that contain structures such as horizons, faults, chaotic layers, and salt domes. We then showed how these image-level labels can be mapped into much more accurate pixel-level labels by solving a simple non-negative matrix factorization problem \cite[]{alaudah2017weakly}\footnote{Both works are better explained in our journal paper currently under review \cite[]{alaudah2018geophysics}.}.

Figure \ref{fig:img_labels} shows examples of the results obtained from our previous work. The figure shows seismic images containing chaotic structures, faults, and salt domes with their corresponding pixel-level labels in blue, green, and red, respectively. While these labels are not as accurate as ones obtained from an expert interpreter, obtaining vast amounts of weak labels, such as these, is rather easy as we've shown in our previous work. Additionally, our method \cite[]{alaudah2017weakly} allows us to easily quantify the confidence in the various pixel-level labels generated for every image. We refer to this confidence in the weak pixel-level labels as $q(x)$, where $x$ is the pixel index.


In addition to our thousands of weakly-labeled images, we further increase the size of our training dataset by using several data augmentation techniques. These techniques include random horizontal flipping, and random rotations of up to $\pm 15^\circ$, of all the images in our dataset and their corresponding labels. This data augmentation step helps to prevent our model from over-fitting to the training data, and helps it generalize to data it hasn't seen before.




\subsection{Deconvolution networks}

\begin{figure*}[ht]

\begin{center}
\includegraphics[width=0.8\textwidth]{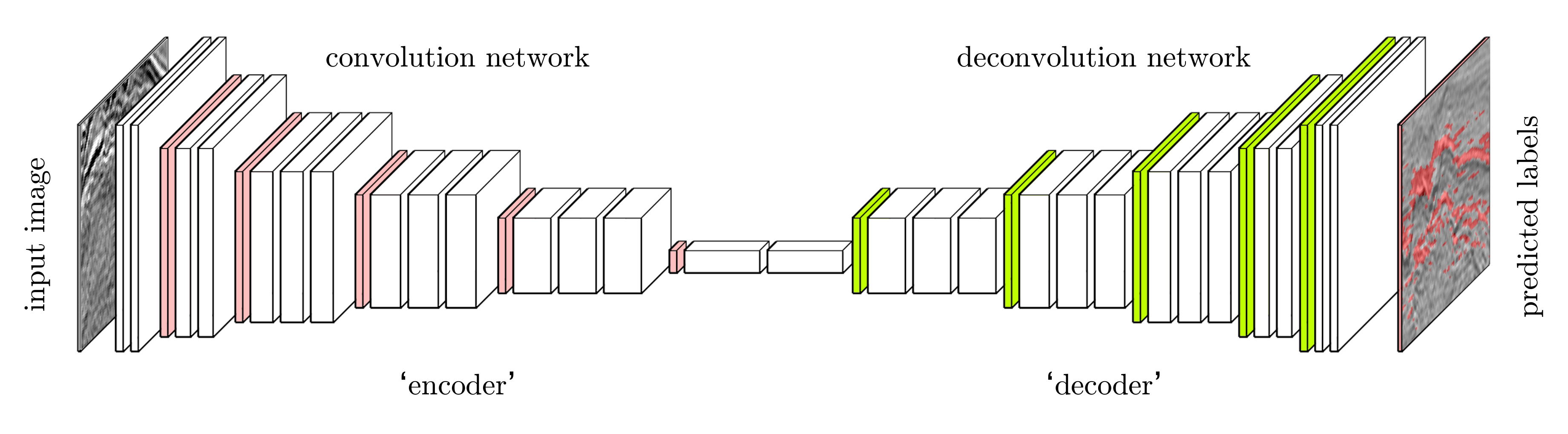}

\end{center}

\caption{The architecture of the deconvolution network used in this work. White layers are convolution or deconvolution layers. Red layers are max-pooling layers, while green layers are unpooling layers.}
\label{fig:deconv}
\end{figure*}

A major hurdle for the successful end-to-end application of CNNs for labeling visual data was what seemed like a trade-off between classification and localization accuracy. Deeper networks that have many convolution and pooling layers have proven to be the most successful models in image classification tasks. However, their large receptive fields and increased spatial invariance (due to pooling and convolutional layers) make it difficult to infer the locations of various objects within the image. In other words, the deeper we go into a network, the more it seemed we lose the location information of various objects within the image. Some researchers have attempted to overcome this hurdle by using various pre- or post-processing techniques. However, the recent introduction of fully convolutional network architectures, such as FCN \cite[]{long2015fully} and DeconvNet \cite[]{noh2015learning} have shown that it is possible to achieve good labeling results using a convolutional network only, with no pre- or post-processing steps required. FCN accomplish this by replacing the fully-connected layers of the CNN with 1D convolutional layers that produce coarse feature maps. These coarse feature maps are then upsampled, and concatenated with the scores from intermediate feature maps in the network to generate the output. These upsampling steps, however, result in a blurred output that loses some of the resolution of the original image. 

Deconvolution networks overcome this problem by using a symmetric encoder-decoder style architecture composed of stacks of convolution and pooling layers in the encoder, and stacks of deconvolution and unpooling layers in the decoder that mirror the encoders architecture. The role of the encoder can be seen as doing object detection and classification, while the decoder is used for accurate localization of these objects. This architecture can achieve finer and more accurate results than those of the FCN, and therefore is adopted in our work. 

Figure \ref{fig:deconv} outlines the architecture of the deconvolution network used in our work. Every convolution or deconvolution layer (in white) is followed by a rectified linear unit (ReLU) non-linearity. The layers in red perform $2\times2$ max pooling to select the maximum filter response within small windows. The indices of the maximum responses for every pooling layer are then shared with their respective unpooling layers (in green) to undo this pooling operation and get a higher resolution image. 

\subsection{Adapting the loss function for learning with weak labels}

\input{other/fig_3D_loss_2.tex}

Since our weak labels are generated automatically, they are not of the same quality as labels obtained from an expert interpreter. However, since obtaining such labels does not require any manual labor nor expensive computational resources, we can use these labels to train our model and modify our network loss function to not trust these weak labels too much. To achieve this, we use a recently introduced loss function called the focal loss \cite[]{lin2017focal} that was recently proposed for dense detection of objects in computer vision tasks. If we write the widely-used cross-entropy loss as
\begin{equation}
    \mathsf{CE}(p(x),q(x)) = -\sum_{x} q(x) \log(p(x)),
\end{equation}
\noindent where $p(x)$ is the confidence of the network output, $q(x)$ is the confidence of the weak labels, and $x$ are the pixels in the image. Then, the focal loss can be written as
\begin{equation}
    \mathsf{FL}(p(x),q(x)) =  \sum_{x} \left(1-p(x)\right)^\gamma \mathsf{CE}(p(x),q(x)),
\end{equation}
\noindent where $\gamma$ is a parameter that controls how much weight is given to regions with low predicted confidence. We use this loss function, as opposed to the more commonly used cross-entropy (CE) loss, to put more weight on misclassified regions in the images and not trust our weak labels as much. Later in the results section, we compare the two loss functions and show how the focal loss can greatly enhance the results when training with weak labels. Figure \ref{fig:fl} shows a comparison of CE with FL for two different values of $\gamma$. As the value of $\gamma$ increases, less emphasis is put on regions where the network has learned the seismic structures relatively well, but not to the degree where they match the weak labels exactly. Instead, more emphasis is put on regions where the network has not learned to classify the underlying structure effectively.     
 
\section{Results}

We train our deconvolution network, shown in Figure \ref{fig:deconv}, on thousands of automatically generated weak labels similar to those shown in Figure \ref{fig:img_labels}. Throughout our training, we set aside $25\%$ of the training data for model selection and validation purposes. Once our model's parameters are selected, we retrain our network on the entirety of the training data.  


\begin{figure}[b]
  \centering
  
  \subfigure[original seismic]
  {\includegraphics[width=1\columnwidth]{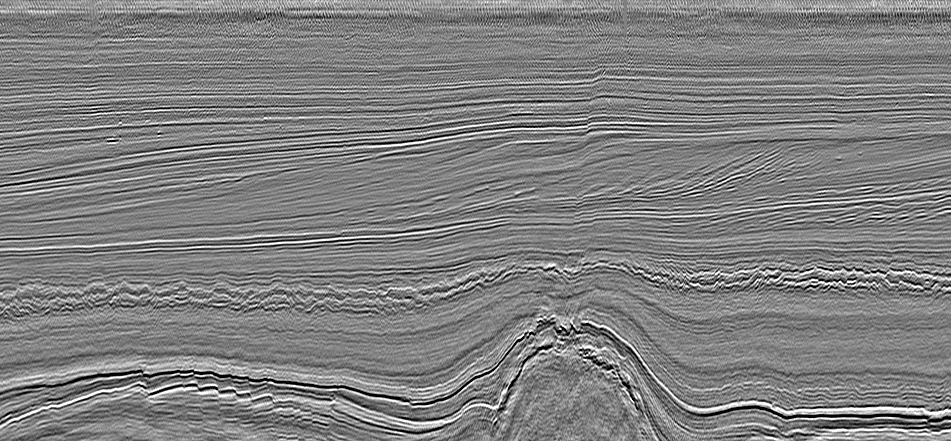}
  }

  \subfigure[\texttt{chaotic} class highlighted]
  {\includegraphics[width=1\columnwidth]{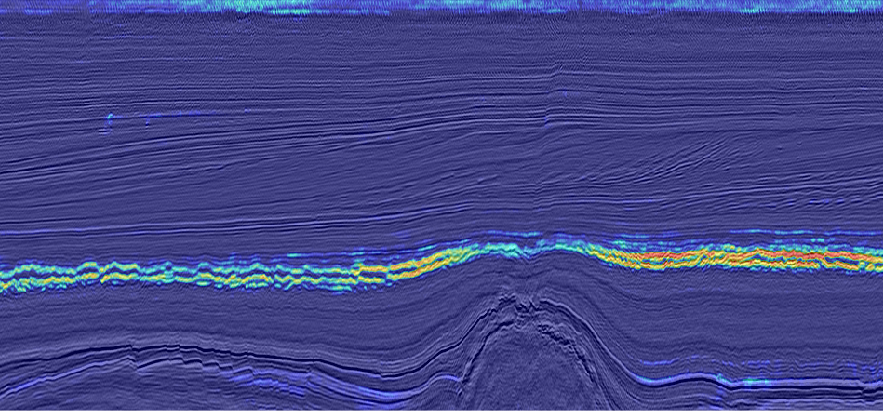}
  }
   \\
   
  \subfigure[\texttt{faults} class highlighted]
  {\includegraphics[width=1\columnwidth]{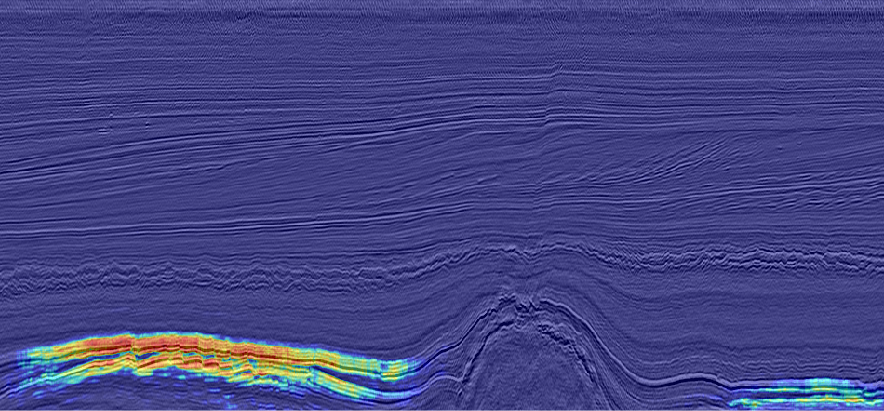}
  }
  
  \subfigure[\texttt{salt dome} class highlighted]
  {\includegraphics[width=1\columnwidth]{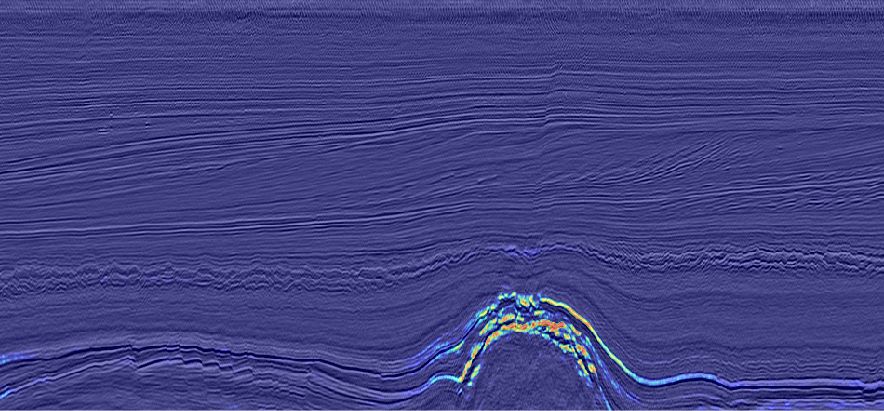}
  }
  
  \caption{Results using our model to highlight various subsurface structures in crossline \#250 of the Netherlands F3 block.}
  \label{fig:main_results}
\end{figure}

\begin{figure}[ht]
  \centering
  \includegraphics[width=1\columnwidth]{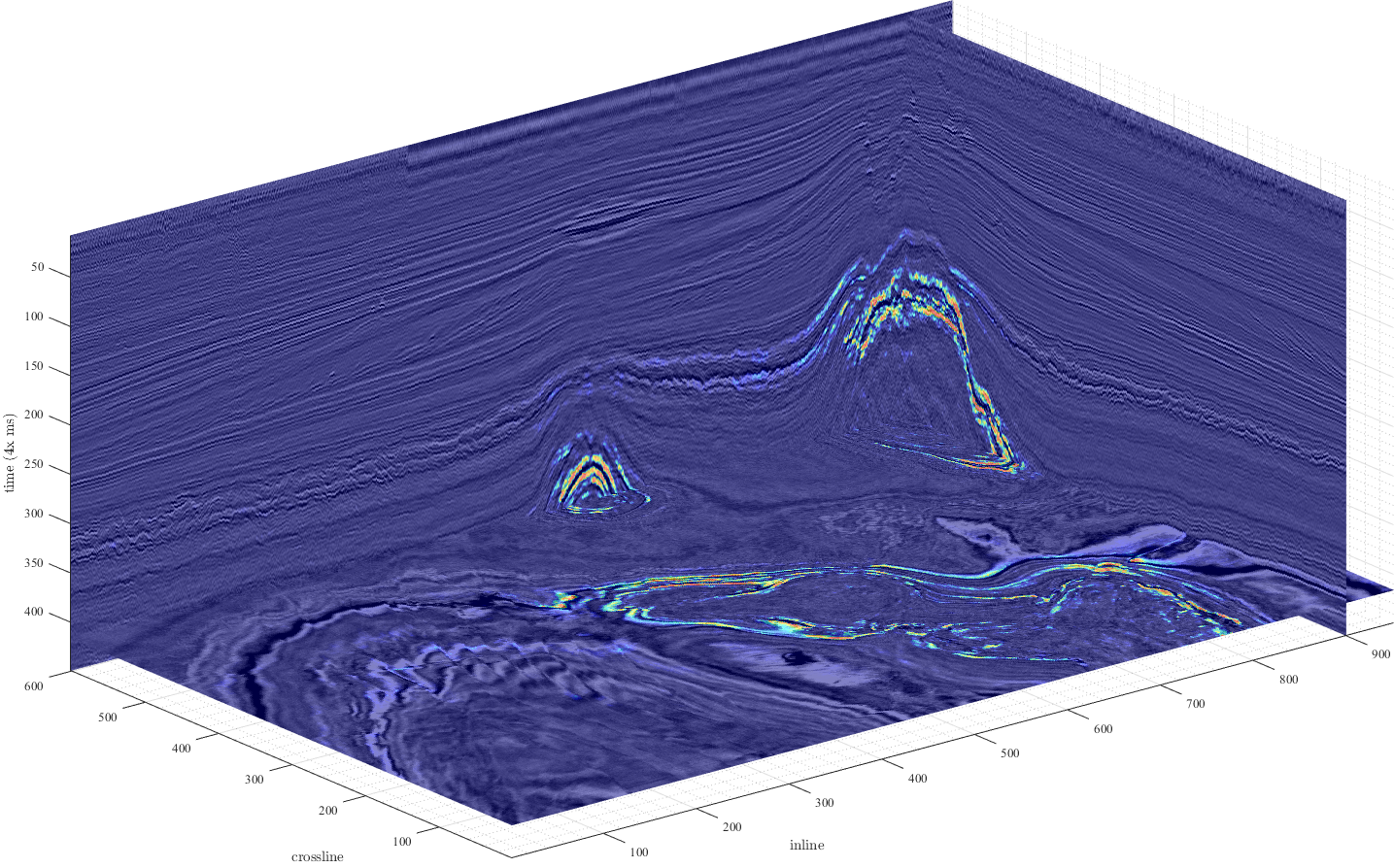}
  \caption{A 3D view of the Netherlands F3 block, with our model highlighting different salt dome structures.}
  \label{fig:3D_salt_dome}
\end{figure}

Once our deconvolution network is trained, we apply it to the Netherlands F3 block \cite[]{F3_data} in a sliding window fashion to label the various subsurface structures in the data. This is done both in the inline and the crossline direction; then the final results are obtained by taking the element-wise product of the two. This step helps reduce any false-positive classifications. Figure \ref{fig:main_results} shows the final results obtained by our model for highlighting \texttt{chaotic}, \texttt{faults}, and \texttt{salt dome} structures in the F3 block. Due to the limited space, only results for crossline \#150 are shown. Additionally, Figure \ref{fig:3D_salt_dome} shows a 3D cross-section of the F3 block with the boundaries of several salt domes highlighted. We note that our model highlights only the salt dome boundaries, and that there are hardly any false positives present in the entire volume. 

\begin{figure}[ht]
  \centering
  
  \subfigure[FCN-8s  using CE]
  {\includegraphics[width=0.70\columnwidth]{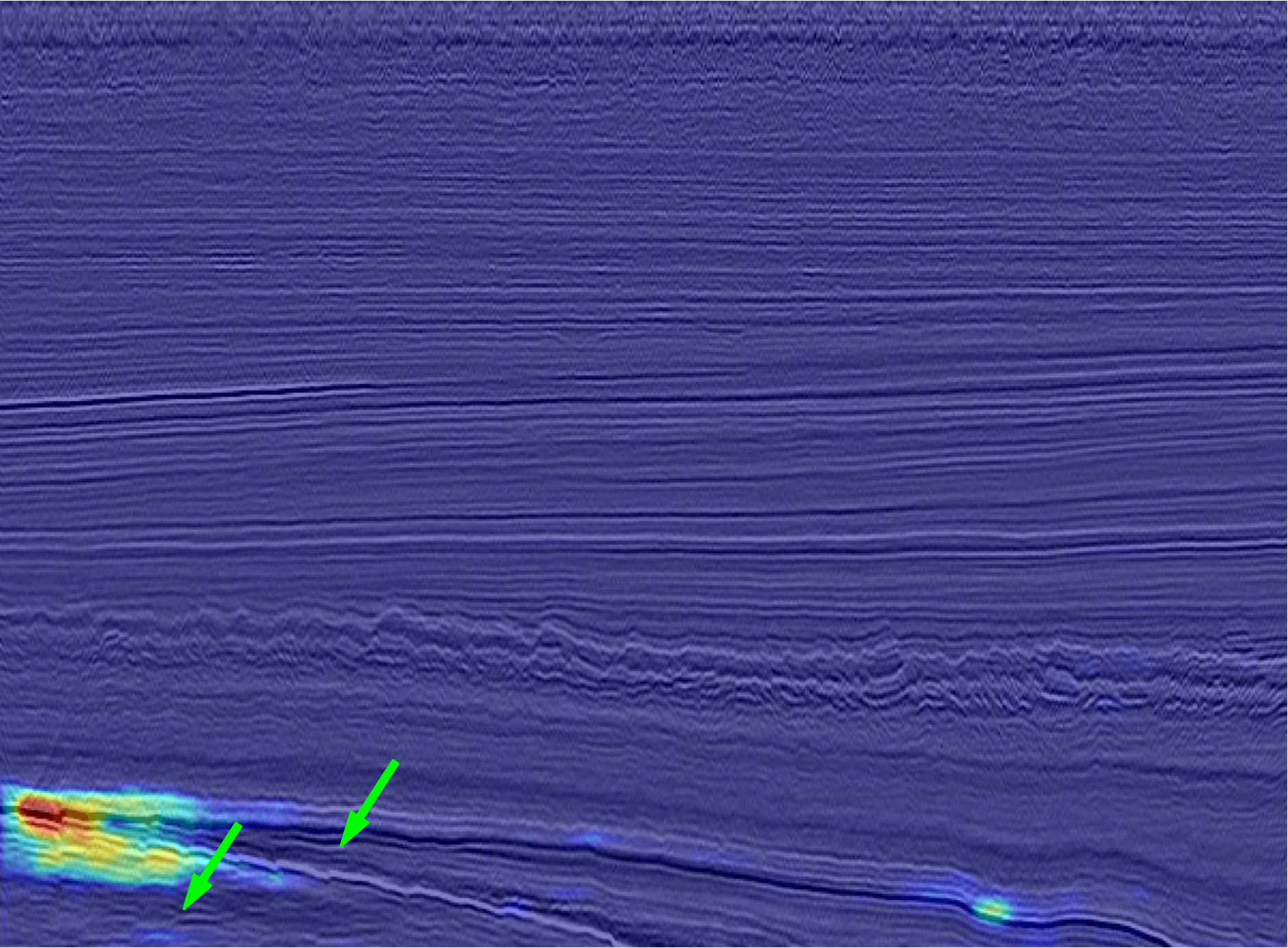}
  }
  
  \subfigure[Deconvolution network using CE]
  {\includegraphics[width=0.70\columnwidth]{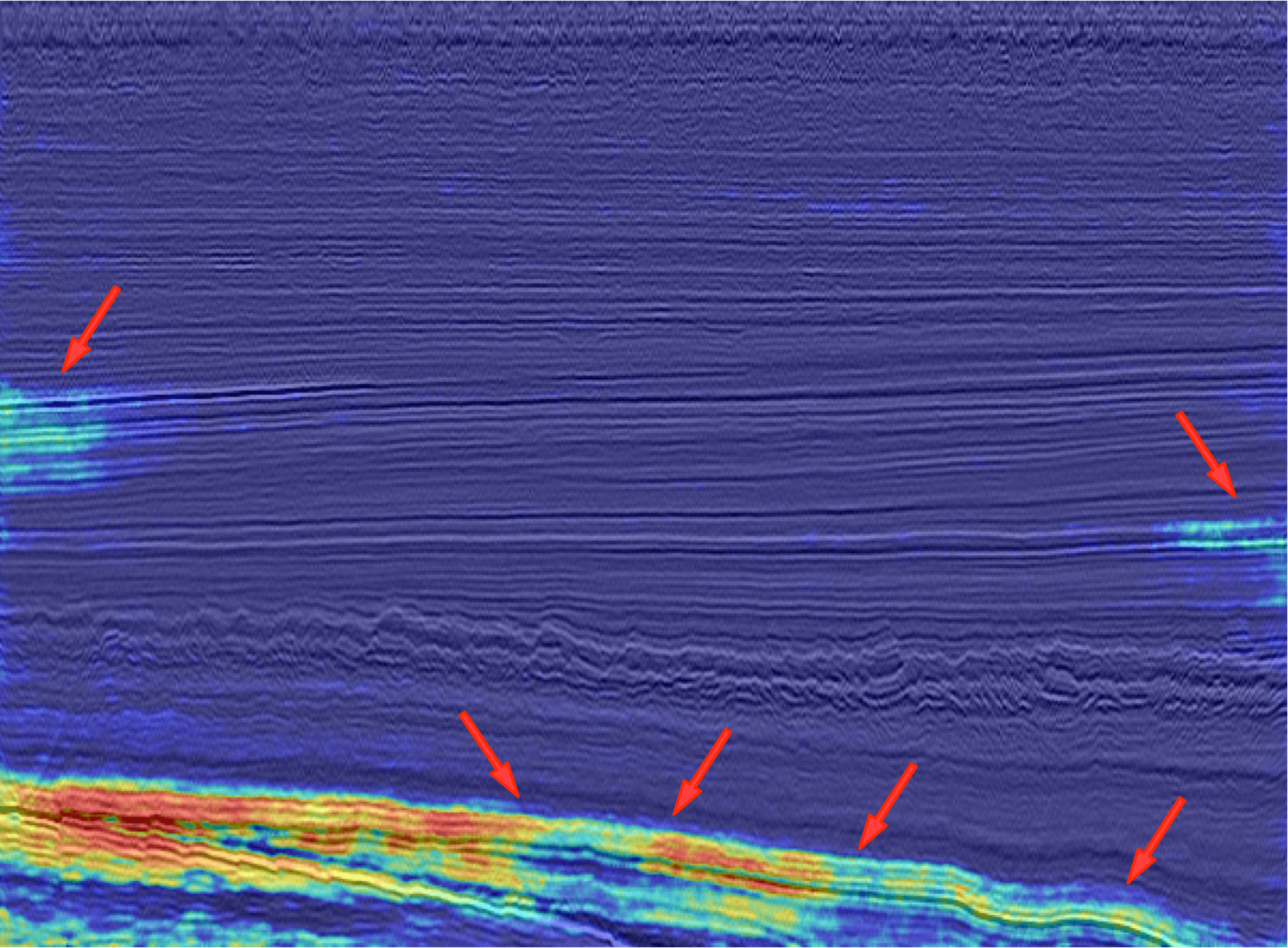}
  }
  
    \subfigure[Deconvolution network using FL (ours)]
  {\includegraphics[width=0.70\columnwidth]{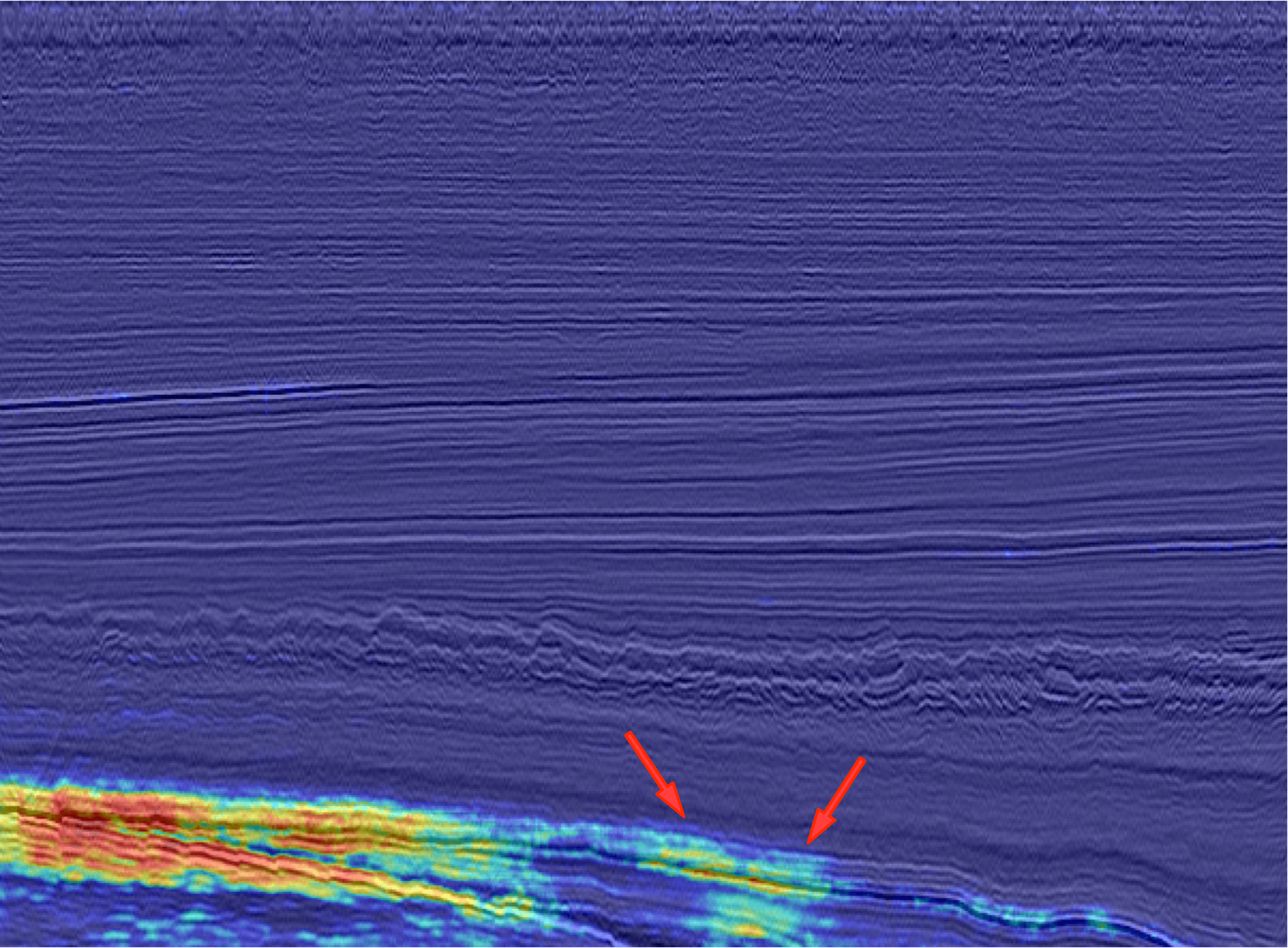}
  }

  \caption{Fault structures in inline \#1 highlighted using either deconvolution network or FCN-8s, and using either the cross entropy loss (CE) or the focal loss (FL). Green arrows indicate false negatives, while red arrows indicate false positives.}
  \label{fig:comparison}
\end{figure}

Additionally, Figure \ref{fig:comparison} shows the results of labeling the \texttt{faults} class in inline \#1 of the F3 block using the FCN-8s network or the deconvolution network with CE or FL losses. FCN-8s is the best performing variant of the FCN architecture proposed by \cite{long2015fully}. We notice that due to the upsampling operation in FCN, several faults in the inline where not labeled. In addition, by comparing (b) and (c) we note how FL helps reduce false positives by not trusting the weak training data too much.  

\section{Conclusions}
In conclusion, we showed how automatically-generated weak labels can be used to train a deep deconvolution network for labeling various subsurface structures within large seismic volumes. We also showed that modifying the loss function of the network to account for the weak labels can help the network avoid false positives, and increase the overall accuracy and robustness of the model.  


\twocolumn

\pagebreak
\bibliographystyle{seg}  
\bibliography{main.bib}

\end{document}